\documentclass[prl,twocolumn,showpacs,superscriptaddress]{revtex4}
\usepackage{amsmath}
\usepackage{amssymb}
\usepackage{graphicx}
\begin{document}

\title{New Evidence for Zero-Temperature Relaxation in a Spin-Polarized Fermi Liquid}
\author{H. Akimoto}
\affiliation{Physics Department, University of Massachusetts, Amherst, MA 01003}
\affiliation{Microkelvin Laboratory, NHMFL, and Physics Department, University of Florida, Gainesville, FL 32611}
\author{D. Candela}
\affiliation{Physics Department, University of Massachusetts, Amherst, MA 01003}
\author{J. S. Xia}
\affiliation{Microkelvin Laboratory, NHMFL, and Physics Department, University of Florida, Gainesville, FL 32611}
\author{W. J. Mullin}
\affiliation{Physics Department, University of Massachusetts, Amherst, MA 01003}
\author{E. D. Adams}
\affiliation{Microkelvin Laboratory, NHMFL, and Physics Department, University of Florida, Gainesville, FL 32611}
\author{N. S. Sullivan}
\affiliation{Microkelvin Laboratory, NHMFL, and Physics Department, University of Florida, Gainesville, FL 32611}

\begin{abstract}
	Spin-echo experiments are reported for $^3$He-$^4$He solutions under extremely high $B/T$ conditions, $B=14.75$~T and $T\geq 1.73$~mK.
	The $^3$He concentration $x_3$ was adjusted close to the value $x_c \approx 3.8\%$ at which the spin rotation parameter $\mu M_0$ vanishes.
	In this way the transverse and longitudinal spin diffusion coefficients $D_\perp$, $D_\parallel$ were measured while keeping $|\mu M_0|<1$.
	It is found that the temperature dependence of $D_\perp$ deviates strongly from $1/T^2$, with anisotropy temperature $T_a = 4.26 \pm 0.18$~mK.
	This value is close to the theoretical prediction for dilute solutions, and suggests that spin current relaxation remains finite as the temperature tends to zero.
\end{abstract}

\pacs{67.65.+z,67.60.Fp,71.10.Ay}
\date{15 September 2002}
\maketitle

	A fundamental result of Fermi liquid theory is that the quasiparticle scattering time and hence the transport coefficients diverge as the temperature tends to zero.
	Recently, there has been much interest in the possibility that spin polarization could remove this divergence for transverse spin currents by creating scattering phase space between spin-up and spin-down Fermi surfaces \cite{meyerovich85, jeon89a, mullin92, meyerovich94, golosov95, golosov98, wei93, ager95, fomin97, vermeulen01, buu02a}.
	Thus the transverse spin diffusion coefficient $D_\perp$ would remain finite at zero temperature in a \emph{partially spin-polarized} Fermi liquid, while the other transport coefficients (longitudinal spin diffusion, viscosity, thermal conductivity) would diverge as in an unpolarized system.
	The \emph{existence} of zero-temperature spin relaxation raises key questions about the applicability of conventional Fermi-liquid theory to transverse spin dynamics, even for weakly polarized systems provided the temperature is sufficiently low \cite{meyerovich94}.

	An initial round of theoretical studies introduced the idea of zero-temperature spin relaxation and computed its magnitude for very dilute systems \cite{meyerovich85, jeon89a, mullin92, meyerovich94, golosov95, golosov98}.
	Spin-echo experiments in polarized liquid $^3$He and $^3$He-$^4$He solutions found large deviations of $D_\perp$ from $T^{-2}$ temperature dependence, supporting the existence of zero-temperature relaxation \cite{wei93, ager95}.
	Experimentally, the apparent magnitude of the effect was stronger than expected from dilute-solution calculations \cite{jeon89a,mullin92}.
	However, the theoretical basis for zero-temperature relaxation has been questioned by Fomin \cite{fomin97}, and a recent experiment based on spin waves rather than spin echoes found that the effect is much weaker than previously measured, if indeed it exists at all \cite{vermeulen01}.
	The origin of the theoretical disagreement is unclear.
	In Ref. \onlinecite{mullin92} a kinetic equation for dilute systems was solved to deduce the existence of zero-temperature attenuation, while in Ref. \onlinecite{meyerovich94} field-theory methods were used to reach a similar conclusion.
	Conversely, in Ref. \onlinecite{fomin97} it was argued that a proper separation of hydrodynamic variables shows that long-wavelength transverse spin currents are not relaxed at zero temperature.

\begin{figure}
\includegraphics[width = 1.0 \linewidth]{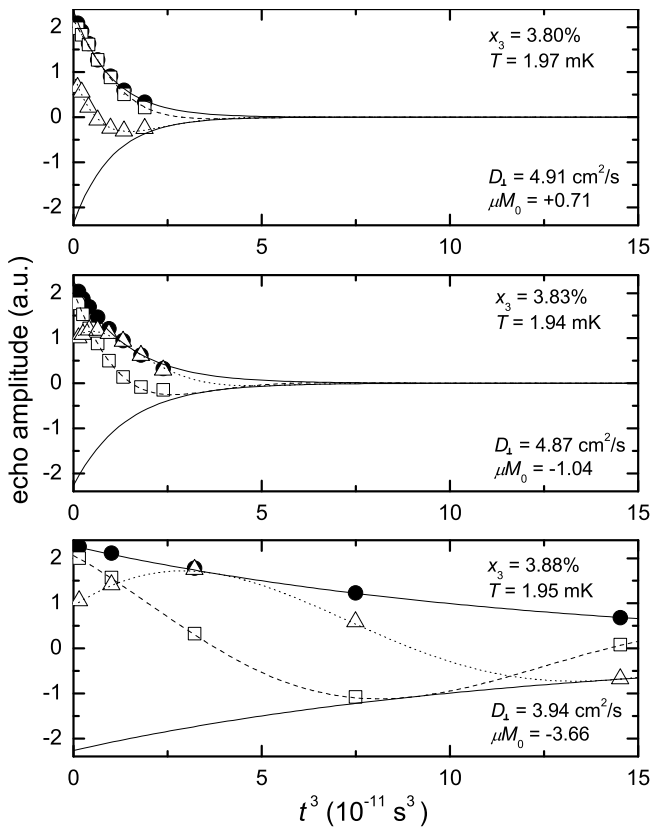}
\caption{\label{figdecay}
	Spin-echo amplitude as a function of time at temperatures near 2~mK for three slightly different values of the $^3$He concentration $x_3$.
	Circles, triangles, and squares show the magnitude, real, and imaginary parts respectively of the measured echo amplitude.
	The curves show fits to Eq. \ref{ht}, which result in the values shown on each graph for the transverse spin diffusion coefficient $D_\perp$ and the spin rotation parameter $\mu M_0$.
	Even though $\mu M_0$ and thus the echo decay time varies rapidly with $x_3$, for all of these data $D_\perp$ is within 12\% of the average value we measure at this temperature, $D_\perp = 4.46$~cm$^2$/s (Table~\ref{tabfits}).
	This should be compared with our measured value of the \emph{longitudinal} diffusion coefficient at the same temperature, $D_\parallel = 16.6$~cm$^2$/s .
}\end{figure}

	In this paper we present new experimental evidence that polarization-induced relaxation does indeed occur in $^3$He-$^4$He, although observation of the effect requires higher fields and/or lower temperatures than previously thought necessary.
	Our results for a 3.8\% $^3$He-$^4$He solution are consistent with the theoretical prediction for extremely dilute solutions, unlike the earlier results.
	Although we use NMR spin echoes, in common with earlier experiments that showed polarization-induced relaxation, we have picked $^3$He concentrations $x_3$ very near the critical concentration $x_c \approx 3.8\%$ at which the spin rotation parameter $\mu M_0 = -\Omega\tau_\perp$ vanishes \cite{ishimoto88,signnote}.
	This is significant because effects that destroy spin echo coherence such as restricted diffusion \cite{buu02} and spin-wave instabilities \cite{ragan00} can limit the apparent magnitude of $\mu M_0$, mimicking a departure from $D_\perp \propto 1/T^2$.
	The earlier experiments were all in the regime $|\mu M_0| >> 1$ apart from one experiment that used a field/temperature ratio $B/T$ ten times lower than that employed in the present work \cite{ager95}.
	By adjusting $x_3$ to within 0.02\% of $x_c$ we achieved the condition $|\mu M_0| < 1$ at our highest $B/T = (14.75\text{~T})/(1.73\text{~mK})$.
	Thus, we have carried out an experiment showing significant spin-diffusion anisotropy ($D_\parallel / D_\perp > 5$) which is robust against possible effects of large $|\mu M_0|$.
	
	The magnitude of polarization-induced relaxation is characterized by an ``anisotropy temperature'' $T_a$ \cite{wei93,meyerovich94}, defined by fitting the transverse diffusion coefficient to 
\begin{equation}\label{Ta}
D_\perp(T) \propto 1/(T^2 + T_a^2).
\end{equation}
	This form implies that $D_\perp$ tends to a constant value as the temperature is lowered well below $T_a$, hence the term ``zero temperature'' relaxation.
	For extremely dilute $^3$He-$^4$He solutions, it is predicted that $T_a = \mu_3 B /2\pi k_B = (248$~$\mu$K/T$)B$ where $\mu_3$ is the $^3$He nuclear magnetic moment \cite{jeon89a,mullin92}.
	Therefore, very high $B/T$ ratios exceeding 4000~T/K are required to measure $T_a$, unless nonequilibrium spin polarization is used as in Ref.~\onlinecite{vermeulen01}.

	To reach these conditions in equilibrium, we have employed a nuclear demagnetization cryostat that incorporates a 15~T NMR-grade sample magnet, which was operated at $B=14.75$~T for the experiments reported here.
	The sample cell has an epoxy NMR tube that extends into but does not touch a 478~MHz NMR resonator \cite{akimoto00} thermally anchored to the mixing chamber.
	With this arrangement we are able to apply 50~W NMR pulses ($180^\circ$ pulse duration = 5.5~$\mu$s) with negligible sample heating.

	The NMR tube consists of a cylindrical sample cavity 2.4~mm dia. $\times$ 2.5~mm high, connected to the main sample cell by a channel 0.7~mm dia. $\times$ 6~mm high.
	The main sample cell contains a sintered heat exchanger (40~m$^2$ area) and three vibrating-wire viscometers.
	One of the viscometers includes an 0.82~mm diameter epoxy rod to reduce slip effects \cite{akimoto02}.
	This viscometer retains temperature sensitivity down to our base temperature of 1.7~mK, and serves as the sample thermometer after calibration at higher temperatures against a $^3$He melting-pressure thermometer outside the high-field region.
	It is important to note that the viscometer directly measures the sample temperature in the main cell with no intervening thermal resistance.
	The sample inside the NMR cavity is in excellent thermal contact with the sample in the main cell (measured time constant $\approx$ 150~ms at the base temperature), due to the $T^{-1}$ temperature dependence of the sample thermal conductivity.
	The mechanical analysis for this composite viscometer and its calibration as a high $B/T$ thermometer will be detailed elsewhere~\cite{akimoto02a}.

	The $^3$He concentration $x_3$ was determined to within $\pm 0.08\%$ (i.e. relative uncertainty of 2\%) by measuring the quantities of gas added to the sample cell.
	To precisely adjust $x_3$ to $x_c$, small quantities of $^4$He were added to the cell between temperature scans.
	Thus, the \emph{differences} between $x_3$ values are known to within $\pm 0.005\%$.

	The transverse spin diffusion coefficient $D_\perp$ and spin-rotation parameter $\mu M_0$ were measured by observing the amplitude $h$ and phase $\phi$ of the spin echoes formed by the two-pulse sequence $\theta$--$t/2$--$180^\circ$--$t/2$--echo.
	Here $\theta = 8^\circ$ is the tipping angle of the first pulse and $t/2$ is the time between pulses.
	To measure $D_\perp$ and $\mu M_0$ at each temperature, a series of spin echo experiments with different delay times $t/2$ were carried out.
	The echo amplitude was fit to the following form, valid for $\theta \ll 90^\circ$ \cite{leggett70}:
\begin{equation}\label{ht}
he^{i\phi} = h_0 \exp[-\mathfrak{D}(\gamma G)^2 t^3/12], \ \ \mathfrak{D}=\frac{D_\perp}{1+i\mu M_0}.
\end{equation}
	The vertical static field gradient $G=29.5 \pm 2.5$~G/cm applied to the sample was accurately measured by a least-squares fit to the shape of a single spin echo.
	The \emph{longitudinal} spin diffusion coefficient $D_\parallel$ was measured by the recovery of the longitudinal magnetization following its modification by a pulse of tip-angle $\theta' = 30^\circ$ and subsequent decay of the transverse components (pulse sequence $30^\circ$--$t$--$8^\circ$)~\cite{nunes92}.

\begin{figure}
\includegraphics[width = 1.0 \linewidth]{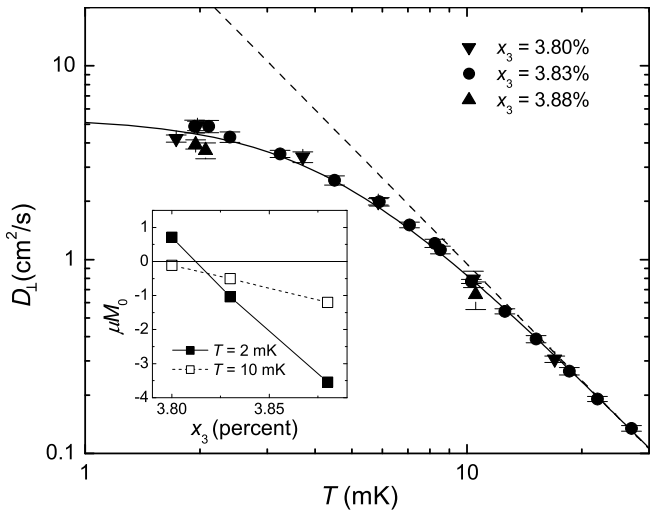}
\caption{\label{figdperp}
	Transverse spin diffusion coefficient $D_\perp$ measured for three values of the $^3$He concentration $x_3$.
	The solid line shows a fit of the data to Eq.~\ref{fitform}, which gives $T_a = 4.26 \pm 0.18$~mK.
	The dashed line shows a $1/T^2$ temperature dependence, corresponding to $T_a = 0$.
	Inset: Fitted spin-rotation parameter $\mu M_0$ as a function of $x_3$ for two temperatures.
}\end{figure}

\begin{figure}
\includegraphics[width = 1.0 \linewidth]{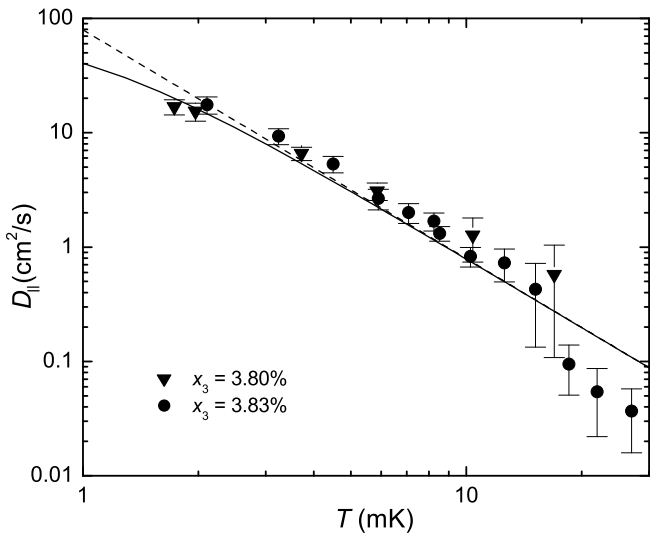}
\caption{\label{figdpar}
	Longitudinal spin diffusion coefficient $D_\parallel$ measured for two of the same $x_3$ values used to measure $D_\perp$.
	The solid and dashed lines show fits as in Fig. \ref{figdperp}.
	For $D_\parallel$, the deviation from $1/T^2$ (difference between solid and dashed lines) is not statistically significant.
}\end{figure}

	Figure \ref{figdecay} shows spin echo data for three closely spaced values of $x_3$, along with fits to Eq.~\ref{ht}.
	Despite the wide variation of $\mu M_0$ and the echo decay time, the fitted $D_\perp$ values agree to within $\pm 12\%$.
	This is further evidence that effects that might limit the apparent magnitude of $\mu M_0$ have not significantly affected the measured value of $D_\perp$.
	Figures \ref{figdperp} and \ref{figdpar} show the data for $D_\perp$ and $D_\parallel$ respectively for several concentrations $x_3$ near the critical concentration $x_c$.
	Table \ref{tabfits} shows the results of fitting these data to
\begin{equation}\label{fitform}
D_{\perp,\parallel}(T) = C_{\perp,\parallel}/(T^2 + T_{\perp,\parallel}^2).
\end{equation}
	Ideally we would find $C_\perp = C_\parallel$, $T_\parallel=0$ and $T_\perp$ would be the experimental estimate of $T_a$.
	In fact, the difference between the fitted values for $C_\perp$ and $C_\parallel$ is within their combined uncertainties, and the fitted value for $T_\parallel$ is consistent with zero.

\begin{table}\caption{\label{tabfits}
	Results of least-squares fits of the NMR data to Eq.~\ref{fitform}.
	In addition to the fit uncertainties listed in this table, there is a 17\% uncertainty in $C_\perp$ due to the uncertainty in the gradient $G$, and a 17\% uncertainty in $C_\parallel$ due to uncertainties in the cell dimensions.
}\begin{ruledtabular}
\begin{tabular}{ccc}
Parameter & Value & Units \\
\hline
$C_\perp$ & $9.79 \pm 0.33 \times 10^{-5}$ & cm$^2$K$^2$/s \\
$T_\perp$ & $4.26 \pm 0.18$ & mK \\
$C_\parallel$ & $7.9 \pm 1.9 \times 10^{-5}$ & cm$^2$K$^2$/s \\
$T_\parallel$ & $0.98 \pm 0.96$ & mK \\
\end{tabular}
\end{ruledtabular}
\end{table}
	
	An important consideration for both $D_\perp$ and $D_\parallel$ measurements is sample heating due to irreversible spin diffusion~\cite{ragan02}.
	We have calculated the temperature rise of a polarized free Fermi gas following a tipping pulse and diffusive decay of the transverse magnetization.
	As in Ref.~\onlinecite{ragan02}, we find that the initial and final temperatures $T_{i,f}$ are related approximately by $T_f^2 = T_i^2 + T_b^2$, and we have calculated $T_b$ numerically as a function of tipping angle and initial polarization and temperature.
	Here $T_b$ is the upper bound for an apparent (false) anisotropy temperature that would be due to this type of heating, in the absence of true spin diffusion anisotropy ($T_a=0$).
	For the conditions of our experiment, we calculate $T_b=1.23$~mK for $\theta=8^\circ$ as used for the $D_\perp$ measurements, and $T_b=3.3$~mK for $\theta'=30^\circ$ as used for the $D_\parallel$ measurements.
	Similar spin-diffusion heating occurs due to imperfections in the $180^\circ$ pulse used to form spin echoes.
	However, we compute that the RMS deviation of the magnetization from a perfect $180^\circ$ rotation is only $9^\circ$ in our experiments, and any such heating would occur well after the main spin echo decay.

	We have checked for several other possible conditions that might affect the $D_\perp$ measurements:
	(1) The smaller of the spin mean free path and the spin rotation distance is never greater than 8\% of the magnetization pitch for an echo decay of $1/e$, consistent with the requirements for the applicability of Leggett's spin dynamic equation~\cite{leggett70}.
	(2) Similarly, we find that for the conditions of our experiment, relaxing the ``steady state'' approximation $\partial\mathbf{J}/\partial t = 0$ used in Ref.~\cite{leggett70} never changes the apparent value of $D_\perp$ by more than a few percent.
	(3) The fitted values of $\mu M_0$ are always much less than the apparent saturation value due to restricted diffusion found in Ref. \cite{buu02}, $(\mu M_0)_{\text{sat}} \approx 0.3 b_L^{1/2}$.
	Here $b_L = L^3(\gamma G) |\mu M_0|/D_\perp$ where $L$ is the cell height.

\begin{figure}
\includegraphics[width = 1.0 \linewidth]{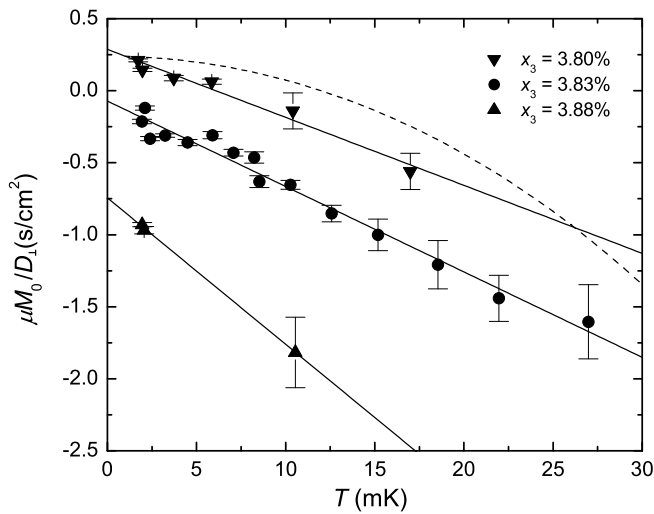}
\caption{\label{figmum}
	Measured ratio of spin rotation parameter $\mu M_0$ to transverse spin diffusion coefficient $D_\perp$ (data points), and linear fits (lines).
	In lowest-order Fermi liquid theory, this ratio should be independent of temperature, depending only upon the Fermi velocity and Fermi-liquid interaction parameters $F_0^a$ and $F_1^a$.
	The dashed line shows the prediction of a calculation based on a phenomenological quasiparticle potential $V(q)$, evaluated for $x_3=3.65\%$.
}\end{figure}

	In Figure.~\ref{figmum} we show our data for the quantity $\mu M_0/D_\perp$.
	In Fermi-liquid theory, $\mu M_0/D_\perp$ is expected to be temperature independent, as both $\mu M_0$ and $D_\perp$ are proportional to the transverse spin current relaxation time $\tau_\perp$ \cite{leggett70}.
	The vanishing of $\mu M_0$ at $x_3 = x_c$ can be viewed as a result of cancellation between positive and negative portions of the quasiparticle interaction potential.
	Therefore, it is perhaps not surprising that this cancellation is upset by thermal excitation of the system, leading to a variation with temperature of $\mu M_0/D_\perp$.
	The observed variation is approximately linear in temperature (Fig.~\ref{figmum}).
	We have checked that the temperature variation is the same on warming and cooling.

	Our most important result is that $D_\perp$ follows Eq.~\ref{Ta} with a nonzero anisotropy temperature $T_a = 4.26 \pm 0.18$~mK, as shown by the solid line fit in Fig.~\ref{figdperp}.
	We believe this result is robust against the effect of spin heating ($T_a > T_b=1.23$~mK, and the latter temperature would only be reached after the echoes fully decayed \cite{ragan02}).
	It is possible that the spin dynamics are modified from Eq.~\ref{ht} at the special point $x_3 = x_c$ \cite{fomin02,meyerovich94}, although we see no experimental evidence of this.
	Interestingly, our measured $T_a$ is nearly consistent with the theoretical value $\mu_3 B /2\pi k_B = 3.73$~mK predicted for very dilute solutions, and is considerably smaller than the value $22 \pm 3$~mK that is obtained by extrapolating earlier results for this $x_3$ to our higher field \cite{ager95}.
	Recently Buu, et al. reanalyzed NMR data taken at $x_3=6.1\%$ taking into account restricted diffusion effects, and concluded that $T_a$ is considerably smaller than previously thought, although still 2.2 times larger than the dilute-solution value \cite{buu02a}.

	From a completely microscopic point of view, it appears impossible at present to predict $T_a$ for $^3$He concentrations outside the s-wave regime, which is roughly $x_3 < 10^{-3}$ \cite{meyerovich85,meyerovich94}.
	To provided some comparison with the present results, we have carried out numerical calculations of $D_\perp$ and $\mu M_0$ using an effective quasiparticle scattering ``potential'' $V(q)$, along the lines of Ref.~\onlinecite{mullin92}.
	Details of these calculations, which must be regarded as semi-phenomenological, will be reported elsewhere \cite{mullin02} and only the results are given here.
	Using the $V(q)$ proposed in Ref.~\cite{baym68} the calculated $\mu M_0$ for $T \rightarrow 0$ crosses zero at $x_3 = 3.65\%$, close to the experimental $x_c$ (Fig.~\ref{figmum}).
	At this $x_3$ (and in fact nearly independent of $x_3$), the calculated anisotropy temperature is $T_a=3.7$~mK.
	Thus, the calculation predicts that the anisotropy temperature retains its s-wave value for $x_3$ far outside the s-wave regime, in agreement with the present experiments.
	Unlike $T_a$ the diffusion coefficients $D_{\parallel,\perp}$ and the spin rotation parameter $\mu M_0$ calculated from $V(q)$ are drastically modified from their s-wave values for this same range of concentrations; they agree well with our experimental values.
	As shown in Fig.~\ref{figmum}, the calculations predict a temperature variation for $\mu M_0/D_\perp$  over the range 2-30 mK that is similar in magnitude to the variation observed experimentally.
	However, the calculated variation is approximately quadratic in temperature, unlike the linear variation seen experimentally.
	This is the only significant discrepancy we find between the experimental results presented here and these calculations based on $V(q)$.

	This work was supported by the in-house research program of the National High Magnetic Field Laboratory funded by NSF DMR-9016241 and the State of Florida.

\bibliography{newev}

\begin{thebibliography}{24}
\expandafter\ifx\csname natexlab\endcsname\relax\def\natexlab#1{#1}\fi
\expandafter\ifx\csname bibnamefont\endcsname\relax
  \def\bibnamefont#1{#1}\fi
\expandafter\ifx\csname bibfnamefont\endcsname\relax
  \def\bibfnamefont#1{#1}\fi
\expandafter\ifx\csname citenamefont\endcsname\relax
  \def\citenamefont#1{#1}\fi
\expandafter\ifx\csname url\endcsname\relax
  \def\url#1{\texttt{#1}}\fi
\expandafter\ifx\csname urlprefix\endcsname\relax\def\urlprefix{URL }\fi
\providecommand{\bibinfo}[2]{#2}
\providecommand{\eprint}[2][]{\url{#2}}

\bibitem[{\citenamefont{Meyerovich}(1985)}]{meyerovich85}
\bibinfo{author}{\bibfnamefont{A.~E.} \bibnamefont{Meyerovich}},
  \bibinfo{journal}{Phys. Lett.} \textbf{\bibinfo{volume}{107A}},
  \bibinfo{pages}{177} (\bibinfo{year}{1985}).

\bibitem[{\citenamefont{Jeon}(1989)}]{jeon89a}
\bibinfo{author}{\bibfnamefont{J.~W.} \bibnamefont{Jeon}}, Ph.D. thesis,
  \bibinfo{school}{University of Massachusetts Amherst} (\bibinfo{year}{1989}),
  \bibinfo{note}{unpublished}.

\bibitem[{\citenamefont{Mullin and Jeon}(1992)}]{mullin92}
\bibinfo{author}{\bibfnamefont{W.~J.} \bibnamefont{Mullin}} \bibnamefont{and}
  \bibinfo{author}{\bibfnamefont{J.~W.} \bibnamefont{Jeon}},
  \bibinfo{journal}{J. Low Temperature Phys.} \textbf{\bibinfo{volume}{88}},
  \bibinfo{pages}{433} (\bibinfo{year}{1992}).

\bibitem[{\citenamefont{Meyerovich and Musaelian}(1994)}]{meyerovich94}
\bibinfo{author}{\bibfnamefont{A.~E.} \bibnamefont{Meyerovich}}
  \bibnamefont{and} \bibinfo{author}{\bibfnamefont{K.~A.}
  \bibnamefont{Musaelian}}, \bibinfo{journal}{Phys. Rev. Lett.}
  \textbf{\bibinfo{volume}{72}}, \bibinfo{pages}{1710} (\bibinfo{year}{1994}).

\bibitem[{\citenamefont{Golosov and Ruckenstein}(1995)}]{golosov95}
\bibinfo{author}{\bibfnamefont{D.~I.} \bibnamefont{Golosov}} \bibnamefont{and}
  \bibinfo{author}{\bibfnamefont{A.~E.} \bibnamefont{Ruckenstein}},
  \bibinfo{journal}{Phys. Rev. Lett.} \textbf{\bibinfo{volume}{74}},
  \bibinfo{pages}{1613} (\bibinfo{year}{1995}).

\bibitem[{\citenamefont{Golosov and Ruckenstein}(1998)}]{golosov98}
\bibinfo{author}{\bibfnamefont{D.~I.} \bibnamefont{Golosov}} \bibnamefont{and}
  \bibinfo{author}{\bibfnamefont{A.~E.} \bibnamefont{Ruckenstein}},
  \bibinfo{journal}{J. Low Temperature Phys.} \textbf{\bibinfo{volume}{112}},
  \bibinfo{pages}{265} (\bibinfo{year}{1998}).

\bibitem[{\citenamefont{Wei et~al.}(1993)\citenamefont{Wei, Kalechofsky, and
  Candela}}]{wei93}
\bibinfo{author}{\bibfnamefont{L.-J.} \bibnamefont{Wei}},
  \bibinfo{author}{\bibfnamefont{N.}~\bibnamefont{Kalechofsky}},
  \bibnamefont{and} \bibinfo{author}{\bibfnamefont{D.}~\bibnamefont{Candela}},
  \bibinfo{journal}{Phys. Rev. Lett.} \textbf{\bibinfo{volume}{71}},
  \bibinfo{pages}{879} (\bibinfo{year}{1993}).

\bibitem[{\citenamefont{Ager et~al.}(1995)\citenamefont{Ager, Child,
  K{\"{o}}nig, Owers-Bradley, and Bowley}}]{ager95}
\bibinfo{author}{\bibfnamefont{J.~H.} \bibnamefont{Ager}},
  \bibinfo{author}{\bibfnamefont{A.}~\bibnamefont{Child}},
  \bibinfo{author}{\bibfnamefont{R.}~\bibnamefont{K{\"{o}}nig}},
  \bibinfo{author}{\bibfnamefont{J.~R.} \bibnamefont{Owers-Bradley}},
  \bibnamefont{and} \bibinfo{author}{\bibfnamefont{R.~M.}
  \bibnamefont{Bowley}}, \bibinfo{journal}{J. Low Temperature Phys.}
  \textbf{\bibinfo{volume}{99}}, \bibinfo{pages}{683} (\bibinfo{year}{1995}).

\bibitem[{\citenamefont{Fomin}(1997)}]{fomin97}
\bibinfo{author}{\bibfnamefont{I.~A.} \bibnamefont{Fomin}},
  \bibinfo{journal}{JETP Lett.} \textbf{\bibinfo{volume}{65}},
  \bibinfo{pages}{749} (\bibinfo{year}{1997}).

\bibitem[{\citenamefont{Vermeulen and Roni}(2001)}]{vermeulen01}
\bibinfo{author}{\bibfnamefont{G.}~\bibnamefont{Vermeulen}} \bibnamefont{and}
  \bibinfo{author}{\bibfnamefont{A.}~\bibnamefont{Roni}},
  \bibinfo{journal}{Phys. Rev. Lett.} \textbf{\bibinfo{volume}{86}},
  \bibinfo{pages}{248} (\bibinfo{year}{2001}).

\bibitem[{\citenamefont{Buu et~al.}(2002{\natexlab{a}})\citenamefont{Buu,
  Clubb, Nyman, Owers-Bradley, and K{\"{o}}nig}}]{buu02a}
\bibinfo{author}{\bibfnamefont{O.}~\bibnamefont{Buu}},
  \bibinfo{author}{\bibfnamefont{D.}~\bibnamefont{Clubb}},
  \bibinfo{author}{\bibfnamefont{R.}~\bibnamefont{Nyman}},
  \bibinfo{author}{\bibfnamefont{J.~R.} \bibnamefont{Owers-Bradley}},
  \bibnamefont{and}
  \bibinfo{author}{\bibfnamefont{R.}~\bibnamefont{K{\"{o}}nig}},
  \bibinfo{journal}{J. Low Temperature Phys.} \textbf{\bibinfo{volume}{128}},
  \bibinfo{pages}{123} (\bibinfo{year}{2002}{\natexlab{a}}).

\bibitem[{\citenamefont{Ishimoto et~al.}(1988)\citenamefont{Ishimoto, Fukuyama,
  Fukuda, Tazaki, and Ogawa}}]{ishimoto88}
\bibinfo{author}{\bibfnamefont{H.}~\bibnamefont{Ishimoto}},
  \bibinfo{author}{\bibfnamefont{H.}~\bibnamefont{Fukuyama}},
  \bibinfo{author}{\bibfnamefont{T.}~\bibnamefont{Fukuda}},
  \bibinfo{author}{\bibfnamefont{T.}~\bibnamefont{Tazaki}}, \bibnamefont{and}
  \bibinfo{author}{\bibfnamefont{S.}~\bibnamefont{Ogawa}},
  \bibinfo{journal}{Phys. Rev. B} \textbf{\bibinfo{volume}{38}},
  \bibinfo{pages}{6422} (\bibinfo{year}{1988}).

\bibitem[{sig()}]{signnote}
\bibinfo{note}{We use the sign convention that $\mu M_0 > 0$ for extremely
  dilute $^3$He-$^4$He.}

\bibitem[{\citenamefont{Buu et~al.}(2002{\natexlab{b}})\citenamefont{Buu,
  Chubb, Nyman, Bowley, and Owers-Bradley}}]{buu02}
\bibinfo{author}{\bibfnamefont{O.}~\bibnamefont{Buu}},
  \bibinfo{author}{\bibfnamefont{D.}~\bibnamefont{Chubb}},
  \bibinfo{author}{\bibfnamefont{R.}~\bibnamefont{Nyman}},
  \bibinfo{author}{\bibfnamefont{R.~M.} \bibnamefont{Bowley}},
  \bibnamefont{and} \bibinfo{author}{\bibfnamefont{J.~R.}
  \bibnamefont{Owers-Bradley}}, \bibinfo{journal}{Phys. Rev. B}
  \textbf{\bibinfo{volume}{65}}, \bibinfo{pages}{13452}
  (\bibinfo{year}{2002}{\natexlab{b}}).

\bibitem[{\citenamefont{Ragan}(2000)}]{ragan00}
\bibinfo{author}{\bibfnamefont{R.~J.} \bibnamefont{Ragan}},
  \bibinfo{journal}{J. Low Temperature Phys.} \textbf{\bibinfo{volume}{121}},
  \bibinfo{pages}{749} (\bibinfo{year}{2000}).

\bibitem[{\citenamefont{Akimoto and Candela}(2000)}]{akimoto00}
\bibinfo{author}{\bibfnamefont{H.}~\bibnamefont{Akimoto}} \bibnamefont{and}
  \bibinfo{author}{\bibfnamefont{D.}~\bibnamefont{Candela}},
  \bibinfo{journal}{J. Low Temperature Phys.} \textbf{\bibinfo{volume}{121}},
  \bibinfo{pages}{791} (\bibinfo{year}{2000}).

\bibitem[{\citenamefont{Akimoto
  et~al.}(2002{\natexlab{a}})\citenamefont{Akimoto, Xia, Adams, Candela,
  Mullin, and Sullivan}}]{akimoto02}
\bibinfo{author}{\bibfnamefont{H.}~\bibnamefont{Akimoto}},
  \bibinfo{author}{\bibfnamefont{J.~S.} \bibnamefont{Xia}},
  \bibinfo{author}{\bibfnamefont{E.~D.} \bibnamefont{Adams}},
  \bibinfo{author}{\bibfnamefont{D.}~\bibnamefont{Candela}},
  \bibinfo{author}{\bibfnamefont{W.~J.} \bibnamefont{Mullin}},
  \bibnamefont{and} \bibinfo{author}{\bibfnamefont{N.~S.}
  \bibnamefont{Sullivan}}, in \emph{\bibinfo{booktitle}{Physical Phenomena at
  High Magnetic Fields IV}}, edited by
  \bibinfo{editor}{\bibfnamefont{G.}~\bibnamefont{Boebinger}}
  \bibnamefont{et~al.} (\bibinfo{publisher}{World Scientific},
  \bibinfo{year}{2002}{\natexlab{a}}).

\bibitem[{\citenamefont{Akimoto et~al.}(2002{\natexlab{b}})}]{akimoto02a}
\bibinfo{author}{\bibfnamefont{H.}~\bibnamefont{Akimoto}} \bibnamefont{et~al.}
  (\bibinfo{year}{2002}{\natexlab{b}}), \bibinfo{note}{unpublished}.

\bibitem[{\citenamefont{Leggett}(1970)}]{leggett70}
\bibinfo{author}{\bibfnamefont{A.~J.} \bibnamefont{Leggett}},
  \bibinfo{journal}{J. Phys. C} \textbf{\bibinfo{volume}{3}},
  \bibinfo{pages}{448} (\bibinfo{year}{1970}).

\bibitem[{\citenamefont{Nunes et~al.}(1992)\citenamefont{Nunes, Jin, Hawthorne,
  Putnam, and Lee}}]{nunes92}
\bibinfo{author}{\bibfnamefont{G.}~\bibnamefont{Nunes}, \bibfnamefont{Jr.}},
  \bibinfo{author}{\bibfnamefont{C.}~\bibnamefont{Jin}},
  \bibinfo{author}{\bibfnamefont{D.~L.} \bibnamefont{Hawthorne}},
  \bibinfo{author}{\bibfnamefont{A.~M.} \bibnamefont{Putnam}},
  \bibnamefont{and} \bibinfo{author}{\bibfnamefont{D.~M.} \bibnamefont{Lee}},
  \bibinfo{journal}{Phys. Rev. B} \textbf{\bibinfo{volume}{46}},
  \bibinfo{pages}{9082} (\bibinfo{year}{1992}).

\bibitem[{\citenamefont{Ragan et~al.}(2002)\citenamefont{Ragan, Grunwald, and
  Glenz}}]{ragan02}
\bibinfo{author}{\bibfnamefont{R.~J.} \bibnamefont{Ragan}},
  \bibinfo{author}{\bibfnamefont{K.}~\bibnamefont{Grunwald}}, \bibnamefont{and}
  \bibinfo{author}{\bibfnamefont{C.}~\bibnamefont{Glenz}}, \bibinfo{journal}{J.
  Low Temperature Phys.} \textbf{\bibinfo{volume}{126}}, \bibinfo{pages}{163}
  (\bibinfo{year}{2002}).

\bibitem[{fom()}]{fomin02}
\bibinfo{note}{I. A. Fomin, private communication (2002)}.

\bibitem[{\citenamefont{Mullin}(2002)}]{mullin02}
\bibinfo{author}{\bibfnamefont{W.~J.} \bibnamefont{Mullin}}
  (\bibinfo{year}{2002}), \bibinfo{note}{unpublished}.

\bibitem[{\citenamefont{Baym and Ebner}(1968)}]{baym68}
\bibinfo{author}{\bibfnamefont{G.}~\bibnamefont{Baym}} \bibnamefont{and}
  \bibinfo{author}{\bibfnamefont{C.}~\bibnamefont{Ebner}},
  \bibinfo{journal}{Phys. Rev.} \textbf{\bibinfo{volume}{170}},
  \bibinfo{pages}{1968} (\bibinfo{year}{1968}).

\end{thebibliography}
\end{document}